\documentclass{jjap3}

\title{Control of Thermoelectric Properties of ZnO using Electric Double Layer}
\author{Ryohei Takayanagi$^{1}$, Takenori Fujii$^{2}$\thanks{E-mail address: fujii@crc.u-tokyo.ac.jp} and Atsushi Asamitsu$^{1,2}$}
\inst{$^{1}$Department of Applied Physics, Univ. of Tokyo, Tokyo 113-8656, Japan \\
$^{2}$Cryogenic Research Center, Univ. of Tokyo, Tokyo 113-0032, Japan}
\abst{We have successfully controlled thermoelectric properties of ZnO by changing carrier concentration using an electric double layer transistor (EDLT) which is a field effect transistor gated by electrolyte solution. The resistivity and the thermopower decreased abruptly by applying gate voltage $V_G$ larger than a threshold voltage ($\sim 2V$), indicating an increase of carrier concentration on the ZnO surface. The temperature dependence of resistivity became metallic, which is characterized by weak temperature dependence of the resistivity, when gate voltage exceeded $2V$. Corresponding to the resistivity, the temperature dependence of thermopower changed remarkably. The thickness of the induced metallic layer was estimated to be about 10 $nm$. And the power factor at $V_G = 4V$ was calculated to $\sim8\times10^{-5}Wm^{-1}K^{-2}$, which is six or seven orders of magnitude larger than that of the bulk substrate ZnO ($V_G = 0$).  Thus EDLT is considered to be a useful way to optimize thermoelectric properties by tuning carrier concentration.}

\begin{document}
\maketitle

\section{Introduction}
Thermoelectric materials are of increasing interest for applications such as power generators and heat pumps. Thermoelectric efficiency is characterized by the figure of merit $Z=S^2/\rho \kappa$, where $S$, $\rho$, and $\kappa$ are thermopower, resistivity, and thermal conductivity, respectively. The key to realize an efficient thermoelectric device lies in finding materials with large power factor $S^2 / \rho$ and low thermal conductivity $\kappa$. It is, however, difficult to control these physical quantities independently since $S$, $\rho$, and $\kappa$ are functions of carrier concentration. According to the thermoelectric transport theories, $ZT$ at optimum carrier concentration is a function of a parameter $m^{\ast 3/2}\mu / \kappa_L$, where $\mu$ is the carrier mobility, $m^{\ast}$ is the 
carrier effective mass and $\kappa_{L}$ is the lattice thermal conductivity. Thus large $m^\ast$ (which increases $S$), large $\mu$ (which decreases $\rho$), and small $\kappa_L$ are required to improve $ZT$. 

On the other hand, low dimensional systems such as thin-film superlattices are known as another way to improve thermoelectric properties. A theoretical prediction indicates that the enhancement of $S$ arises from an increase in the density of states near the edge of conduction band when electrons are confined in quantum-well structure\cite{QW1}. Such an enhancement of $S$ has been confirmed experimentally in PbTe/Pb$_{0.927}$Eu$_{0.073}$Te superlattice\cite{QW2}. In the case of transition metal oxides, the improvement of thermoelectric propereties have been reported in SrTiO$_3$/SrTi$_{0.8}$Nb$_{0.2}$O$_3$ superlattice\cite{STO1} and in the {\it water electrolyte} gated field effect transistor (FET) of SrTiO$_3$\cite{STO2}.
Recently, it was found that a two-dimensional electron gas (2DEG) in MgZnO/ZnO heterointerface, where the mismatch in spontaneous polarization of two piezoelectric compounds is compensated by the charge accumulation at the interface, shows a very large mobility over 300,000 $cm^2 / Vs$\cite{TFZnO1,TFZnO2}. Therefore, 2DEG on ZnO is considered to be promising for the thermoelectric application. Moreover, bulk ZnO is known as a n-type thermoelectric material which shows large power factor of $\sim 2 \times 10^{-3}Wm^{-1}K^{-2}$, and there have already been a number of studies to optimize thermoelectric properties by doping chemically\cite{ZnO1,ZnO2,ZnO3}.
Here, we controlled the carrier concentration by using electric double layer transistor (EDLT) on ZnO surface and measured its thermoelectric properties. EDLT is a field effect transistor (FET) whose insulating layer is replaced with polymer electrolytes or ionic liquids. When a gate voltage ($V_G$) is applied between the gate electrode and the sample, ions in the liquid are aligned along the surface of the sample, forming a charged double layer which is a kind of nanoscale capacitor. This technique has the advantage of accumulating carriers up to $\sim10^{15} cm^{-2}$ on the interface of the sample, which is one or two orders of magnitude larger than that in conventional FETs. The capability of accumulating large number of carriers have given rise to new findings such as metal-insulator phase transition (MIT) in ZnO\cite{EDLT1,EDLT2} and electric-field induced superconductivity in SrTiO$_3$\cite{EDLT3}, ZrNCl\cite{EDLT4}, KTaO$_3$\cite{EDLT5} and MoS$_2$\cite{EDLT6}. In this paper, we report first achievement of controlling the thermoelectric properties of ZnO by using EDLT.

\section{Experiment}
ZnO polycrystalline sample was synthesized by a conventional solid-state reaction method. A commercial ZnO powder (4N) was pressed into a pellet and sintered at 1000$^{\circ}C$ for 15h in air. The sintered sample was cut in rectangular shape (about 3 $\times$ 1 $\times$ 0.3 $mm^3$) and polished with an abrasive of one-micron aluminum-oxide (3M Lapping Film) to make the surface flat. A commercial single crystal was also used for EDLT in order to confirm the effects of the surface condition on the EDLT operation. Schematic pictures of EDLT are shown in Fig. 1. The source and drain electrodes were made by Au paste and covered with silicon adhesive sealant to avoid chemical reaction between electrolytes and Au electrodes. 
The gate electrode was made of Pt foil in order to avoid chemical reactions. As seen in the Fig. 1(a) and (b), the resistivity and the thermopower were measured simultaneously with the same configuration. When a positive $V_G$ was applied between the gate and source electrodes, cations in the elecctrolyte were aligned along the ZnO surface, and a negative image charge was induced in the surface of ZnO sample. We chose the electrolyte of KClO$_4$, which was solvated by polyethylene glycol (PEG), ([K]:[O] in PEG = 1:20). The melting point of PEG (molecular weight 1,000) is 310$K$, and ions (K$^+$ and ClO$^{4-}$) in gelatinous PEG can move above 260$K$. In this measurement, $V_G$ was always changed at 300$K$, since electrochemical reaction between ZnO surface and electrolyte occurs at a high temperature.
As shown in Fig. 1(a), the resistivity $\rho$ was measured by two-probe method where the drain current ($I_D$) was monitored with applying the drain voltage ($V_D$) of 0.1$V$. Fig. 1(b) shows the setup for the measurement of thermopower. A temperature gradient was applied with a heater attached near the source electrode (the hot side of the sample) and the temperature difference $\Delta T$ was monitored by using cupper-constantan differential thermocouple. Since the leak current between Pt gate and the sample induces the voltage drop ($V_0 \sim I_G \times R = (I_G \times V_D) /I_D $), an offset voltage $V_0$ was added in the measured $\Delta V$. However if $V_0$ was carefully subtracted from $\Delta V$, the actual thermal electromotive force ($\Delta V$ - $V_0$) was proportional to $\Delta T$ at each $V_G$. At the temperatures below 260K, the leak current was significantly small, because PEG was completely frozen. There, no offset voltage was observed and the measured $\Delta V$ was perfectly linear to $\Delta T$. Hence we can accurately determine the value of S from the slope of thermal electromotive force. Here, $\Delta V$ between source and drain electrodes was measured at two different $\Delta T$ around 1$K$ and 2$K$, and the thermopower was determined by $S = (\Delta V_2-\Delta V_1)/(\Delta T_2-\Delta T_1)$. To measure Hall-coefficient, Hall electrodes were attached perpendicular to the drain current direction. Hall voltage ($V_H$) was measured by constant current method ($I_D = 1 \mu A$) with sweeping magnetic field between $\pm 5T$.
All measurements were carried out under an atmosphere of helium gas so as to prevent the electrolyte from absorbing water, which induces the gate leak current ($I_G$) between Pt gate and the sample. We have confirmed that the thermal conductivity of PEG was about two orders of magnitude lower than that of polycrystalline ZnO. Thus it is considered that EDLT device structure hardly affect the thermal properties of the sample.

\section{Result and Discussion}

Fig. 2(a) shows the temperature dependence of the resistivity for polycrystalline ZnO annealed in oxygen, air, and argon atmosphere. Oxygen defects which naturally built in ZnO decreases with oxygen partial pressure of annealing. As seen in fig. 3(a), the absolute value of the resistivity at room temperature decreases with decreasing oxygen contents (increasing defects) which induce electron doping. The resistivity of single crystal is also plotted for comparison. It shows lowest resistivity at room temperature. This is thought to due to large carrier concentration and/or large mobility compared to polycrystalline sample. 
Fig. 2(b) shows the transfer curve (drain current $I_D$ versus gate voltage $V_G$) of polycrystalline samples and single crystal, which measured at a drain voltage $V_D$ of 0.1 $V$ and a temperature of 300 $K$. The measurement of $I_D$ was carried out with changing $V_G$ from 0 $V$ to 4 $V$, and back to 0 $V$. The sweep rate was as slow as 0.05 $V / min$, which prevents a hysteresis caused by slow response of electrolyte to the applied electric field. It has been confirmed that $I_D$ reproduce the initial values after gate sweeping. Therefore the carriers were induced reversibly and no electrochemical reactions occurred on the surface of the sample. $I_D$ increased abruptly by applying $V_G$ larger than a threshold voltage (about 2 $V$), and leveled off above $V_G = 3 V$. Thus these EDLT devices exhibited typical transistor behavior although the surface of the polycrystalline ZnO was not atomically flat. It is thought that an atomically flat surface is required for the EDLT operation, because the ions in the liquid form a nanoscale capacitor on the surface. However, as seen in the figure, EDLT operation is not sensitive to the surface, and it has been observed both in single crystal and polycrystalline sample. Rather, the on-off ratio of the device in polycrystalline sample was larger than that of single crystal. Since the off-state current is determined by a range of bulk resistivity, the polycrystalline sample annealed in oxygen showed largest on-off ratio. Hereafter, we discuss polycrystalline sample annealed in Air, since it shows not only large on-off ratio but also large conductivity in the on-state. 

The gate voltage dependences of $I_D$, $I_G$, and $S$ at room temperature are shown in Fig. 3. As seen in Fig. 3(b), the leak current of this device was quite low, which was comparable in magnitude to that reported in ZnO single crystals\cite{EDLT1}. In Fig. 3(c), a sharp decrease of $S$ was observed at the same threshold, implying that the surface of ZnO becomes metallic by carrier doping. In the FET configuration, metallic surface and insulating bulk substrate are considered to form a parallel circuit. Therefore, the observed total sheet conductivity ($\sigma_{\Box total}$) and thermopower ($S_{total}$) is given by 
\begin{displaymath}
\sigma_{\Box total}=\sigma_{\Box sur}+\sigma_{\Box bulk}
\end{displaymath}
\begin{displaymath}
S_{total}=\frac{S_{sur}\sigma_{\Box sur}+S_{bulk}\sigma_{\Box bulk}}{\sigma_{\Box sur}+\sigma_{\Box bulk}}
\end{displaymath}
where $S_{sur}$ and $\sigma_{\Box sur}$ are the thermopower and the sheet conductivity of the metallic surface, and $S_{bulk}$ and $\sigma_{\Box bulk}$ are those of insulating bulk substrate, respectively. As discussed later, since the thickness of the surface metallic layer is five order of  magnitude smaller than that of bulk substrate, we considered $\sigma_{\Box bulk}$ as conductivity at $V_G$ = 0 $V$. Fig 4(a) shows $\sigma_{\Box total}$ and $\sigma_{\Box sur}$ as a function of $V_G$. Using $\sigma_{\Box sur}$ and $S_{bulk}$ (thermopower at $V_G$ = 0), $S_{sur}$ was calculated in Fig. 4(b). As seen in the Fig. 4, $\sigma_{\Box total}$ and $S_{total}$ is dominated by those of the metallic surface at large $V_G$.
Thus, these results strongly prove that we can measure the transport properties of the metallic surface layer above $V_G$ = 2$V$.  

In the case of SrTiO$_3$/SrTi$_{0.8}$Nb$_{0.2}$O$_3$ superlattice, a drastic increase in $S$ at room temperature ($S_{300K}$) is observed when the thickness of the SrTi$_{0.8}$Nb$_{0.2}$O$_3$ layer becomes less than $1.56nm$ (corresponding to a four unit cell layer thickness)\cite{STO1}. The $S_{300K}$ value for the one unit cell layer thickness of the SrTi$_{0.8}$Nb$_{0.2}$O$_3$ reached $480 \mu VK^{-1}$, which is 4.4 times larger than that of the 3D bulk SrTi$_{0.8}$Nb$_{0.2}$O$_3$. 
On the other hand, in the case of EDLT of SrTiO$_3$\cite{EDLT3}, it is reported that an increase of $V_G$ reduces the thickness of the accumulation layer on SrTiO$_3$, because an increase of $V_G$ reduces the dielectric constant of SrTiO$_3$ owing to its incipient ferroelectricity. Therefore, drastic increase in $S$ is expected when the $V_G$ increases to be shrank the accumulation layer. However, as seen in Fig. 3(c), enhancement of $S$ was not observed at large $V_G$, but the thermopower and the resistivity decrease simultaneously like the behavior of the conventional doping effect.

To investigate whether the thermoelectric properties of accumulation layer are two-dimensional or not, the Jonker-plot, in which $S$ was plotted with respect to the sheet conductivity ($\sigma_{\Box}$), is shown in Fig. 5. The relation of $S$ and $\sigma$ is theoretically explained\cite{STO1,jonker} by
\begin{displaymath}
S=-a \log\sigma +b
\end{displaymath}
When the system has 3D energy band with a parabolic density of states near the Fermi surface, parameter $a$, that is the slope of the $S-\log \sigma$ line, is displayed as
\begin{displaymath}
a=\frac{k}{e} \ln 10=198 \mu V/K
\end{displaymath}
As seen in Fig. 5, the value of $S$ decrease with increasing $\sigma_{\Box}$, and the slope of $S$ is in good agreement with the theoretical prediction for 3D system ($-198\mu V/K$), which is shown by the triangle in Fig. 5. These results indicate that the thickness of the surface metallic layer is not as thin as STO superlattice, as discussed later.

Fig. 6(a) shows the temperature dependence of sheet resistivity ($\rho_{\Box}$) at various $V_G$. $\rho_{\Box}$ at low $V_G$ (below 1$V$) showed semiconducting behavior, in which the resistivity increased abruptly at low temperature. In contrast, at high $V_G$ (above $2V$), the temperature dependence of $\rho_{\Box}$ was weak and the absolute value of $\rho_{\Box}$ was significantly small particularly at low temperatures, indicating that a large amount of electrons were accumulated on the surface and that a metallic state was realized. The temperature dependence of $S$ at various $V_G$ were shown in Fig. 6(b). Corresponding to the resistivity, the temperature dependence of $S$ changed remarkably. Below the threshold voltage, the resistivity was too high to measure thermopower at low temperatures. At high $V_G$, the absolute value of $S$ decreased with lowering temperature, and was nearly constant below 200$K$. These behaviors of $S$ are considered to due to the metallic state of the surface. 

In order to estimate the carrier concentration ($n$), Hall-effect measurement was carried out. Here, we used the sample different from measurement of resistivity and thermopower. Although the thickness of the surface metallic layer ($d$), and hence the $n$, cannot be directly obtained, we can roughly estimate $d$ by using the sheet carrier concentration $n_{\Box}$ (evaluated from Hall coefficient) and the critical carrier concentration ($n_c$) at Metal-Insulator transition (MIT) in ZnO.  In general, a metallic state emerges when the mean distance of donors $r$, which is deduced from donor density (carrier concentration) $n=(4/3\pi r^3)^{-1}$, becomes much smaller than the effective Bohr radius ($a_B$), which is 1.7 $nm$ for ZnO. Therefore, the $n_c$ is calculated as $n_c=3/4\pi a_B^3\sim$ 4.9$\times$10$^{19} cm^{-3}$\cite{Hall}. Fig. 6(c) shows $n_{\Box}$ obtained by Hall-effect measurement. The temperature dependence of resistivity showed metallic behavior when $n_{\Box}$ exceeded 5$\times$10$^{13} cm^{-2}$ (shown as dotted line in Fig. 4(c)). At the critical point (MIT), the formula $n_c=n_{\Box}/d$=5$\times$10$^{13}/d$ should be established. Thus, $d$ was estimated to be approximately 10$nm$, which is consistent with the typical thickness of carrier accumulating layer for EDLT\cite{EDLT3}. The reason why an enhancement of $S$ was not observed in this system, is probably because the thickness of 2DEG in our system (10$nm$) is too large to emerge quantum confinement of electrons. Using the obtained $d$, the power factor of this EDLT device with applying $V_G$ = 4$V$ is estimated to be 8$\times$10$^{-5}Wm^{-1}K^{-2}$ at 300$K$. This value is six or seven orders of magnitude larger than that of the bulk state ($V_G$ = 0). Although the power factor is not as large as reported ZnO ceramics of optimum doping condition ($\sim$ 2$\times$10$^{-3}Wm^{-1}K^{-2}$), EDLT is considered to be a useful way to optimize thermoelectric properties by tuning carrier concentration.

\section{Conclusions}
We have successfully controlled the thermoelectric properties by using EDLT configuration in ZnO. The resistivity and the thermopower decreased drastically by applying a gate voltage larger than a threshold voltage, indicating the increase of carrier concentration at the surface. We have confirmed the reproducibility of gate-voltage dependence for the resistivity and the thermopower. This fact suggests that the carriers are induced reversibly and that no electrochemical reactions occurred on the surface. Contrary to one's expectation, these observations appeared both in single crystal and powder sample implying the EDLT function is not robust in surface condition although the carriers are induced in the vicinity of the surface. Three dimensional behavior have been observed in the Jonker-plot (Slope of the $S$-log$\sigma$ plot), which indicates the metallic surface layer is not enough thin compared to SrTiO$_3$/SrTi$_{0.8}$Nb$_{0.2}$O$_3$ superlattice. The power factor was estimated as 8$\times$10$^{-5}Wm^{-1}K^{-2}$ at 300$K$, which is six or seven orders of magnitude larger than that of the bulk ZnO ($V_G$ = 0). Thus we consider that the thermoelectric properties are improved by controlling carrier concentration by using EDLT configuration in ZnO.

\newpage

\begin{figure}
\begin{center}
\includegraphics[width=12cm]{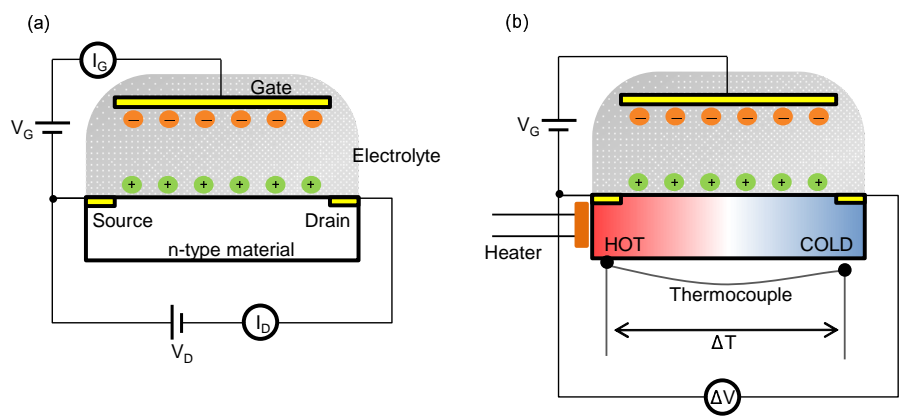}
\end{center}
\caption{Schematic figures of electric double layer transistor. (a) Configuration for measuring resistivity. (b) Configuration for measuring thermopower.}
\label{f1}
\end{figure}

\begin{figure}
\begin{center}
\includegraphics[width=10cm]{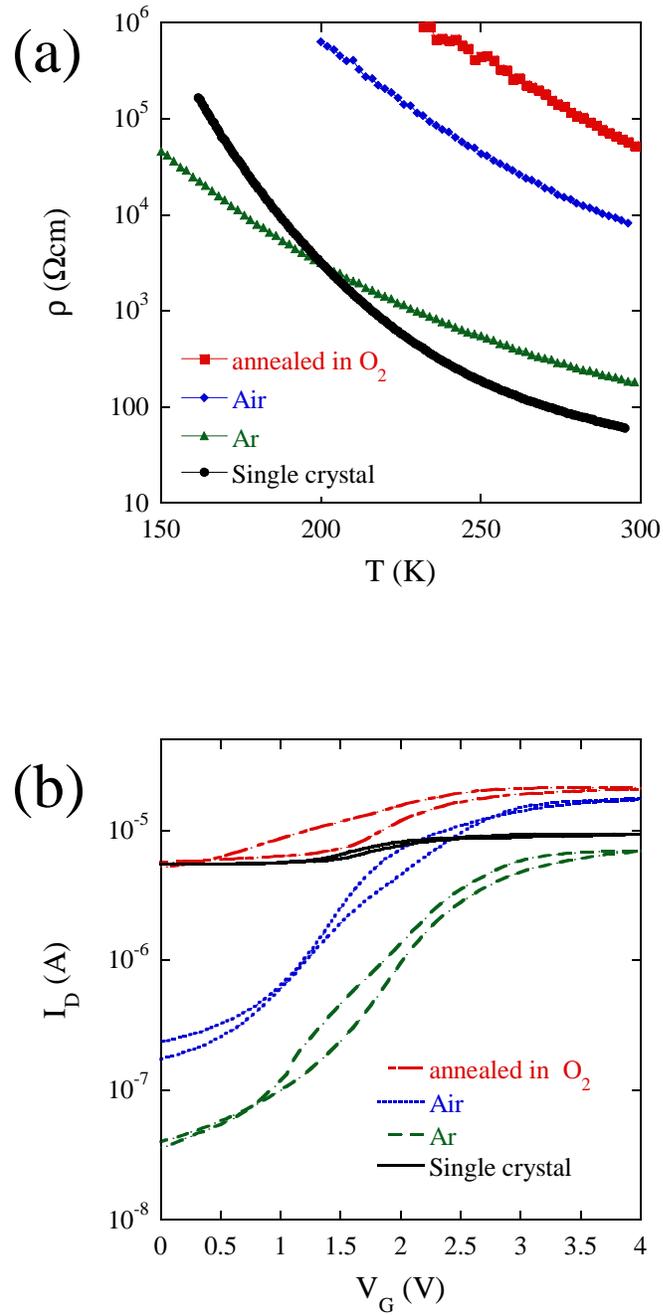}
\end{center}
\caption{(a): Temperature dependence of the resistivity for polycrystalline ZnO annealed in various atmosphere and single crystal. (b): Transfer curve (drain current $I_D$ versus gate voltage $V_G$) of polycrystalline samples and single crystal. }
\label{f2}
\end{figure}

\begin{figure}
\begin{center}
\includegraphics[width=10cm]{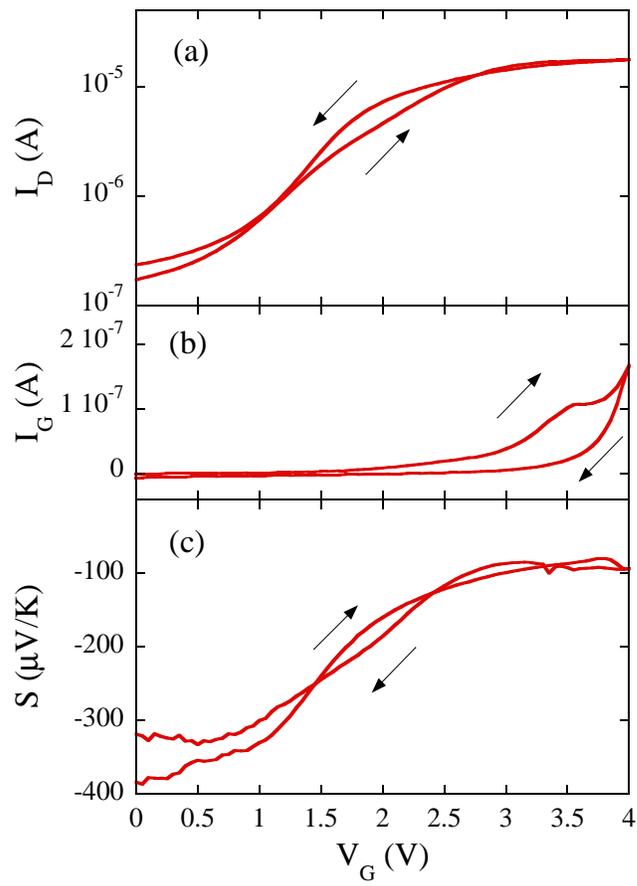}
\end{center}
\caption{Gate voltage ($V_G$) dependence of $I_D$, $I_G$, and $S$ at room temperature. }
\label{f3}
\end{figure}

\begin{figure}
\begin{center}
\includegraphics[width=10cm]{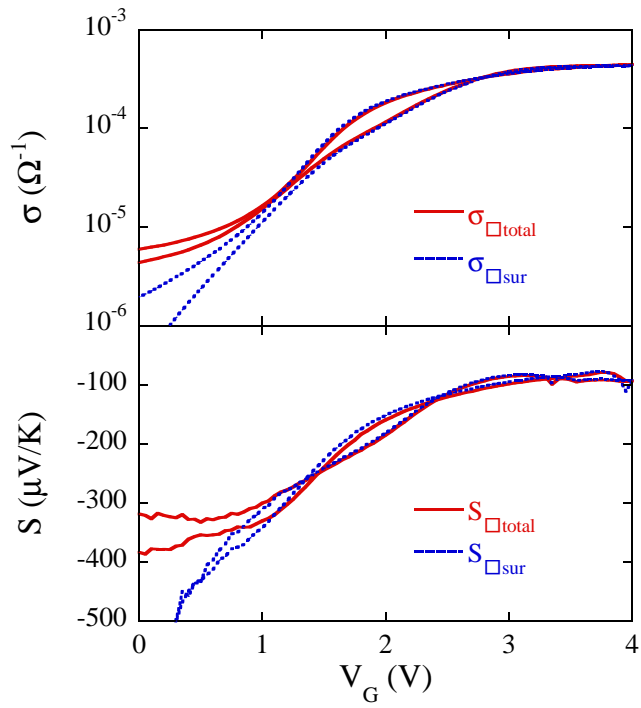}
\end{center}
\caption{(a): $\sigma_{\Box total}$ and $\sigma_{\Box sur}$ are plotted as a function of $V_G$. Here, we considered $\sigma_{\Box bulk}$ as conductivity at $V_G$ = 0.  (b): $S_{total}$ and $S_{sur}$ versus $V_G$. $S_{sur}$ was calculated by using $\sigma_{\Box sur}$ and $S_{bulk}$ (thermopower at $V_G$ = 0). }
\label{f4}
\end{figure}

\begin{figure}
\begin{center}
\includegraphics[width=10cm]{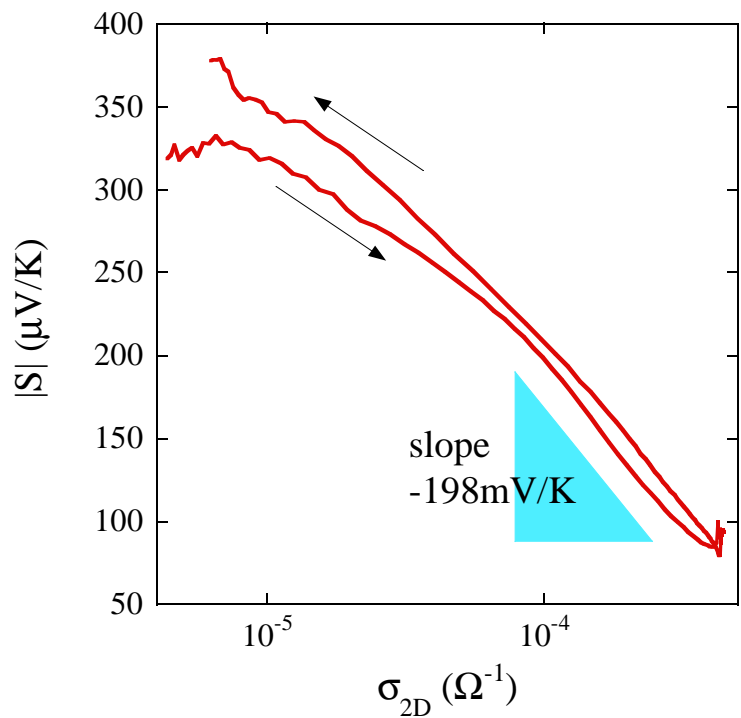}
\end{center}
\caption{Absolute value of S as a function of sheet conductivity. (Jonker plot)}
\label{f5}
\end{figure}

\begin{figure}
\begin{center}
\includegraphics[width=10cm]{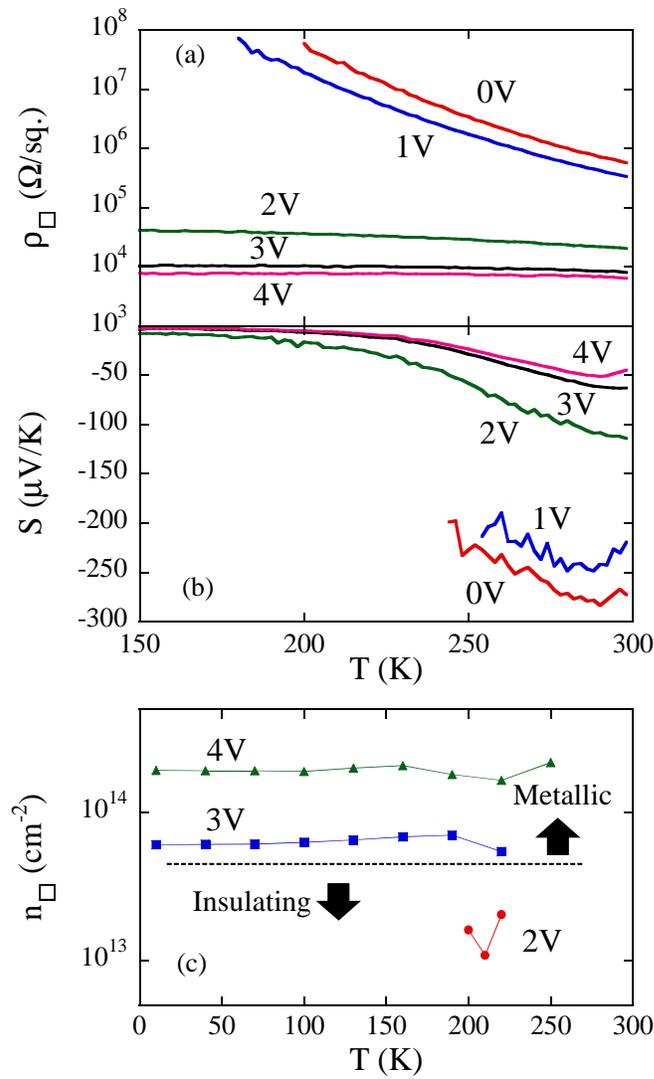}
\end{center}
\caption{Temperature dependence of (a) sheet resistivity, (b) Seebeck coefficient and (c)sheet carrier concentration at various $V_G$.}
\label{f6}
\end{figure}

\end{document}